\documentclass[preprint,aps]{revtex4}

\usepackage{graphicx}

\begin{document}


\title{Electronic crossover suggested by Raman scattering in overdoped (Y,Ca)Ba$_2$Cu$_3$O$_y$}

\author{T. Masui}
\email{tmasui@phys.sci.osaka-u.ac.jp}
\author{T. Hiramachi}
\author{K. Nagasao}
\author{S. Tajima} 
\affiliation{Dept. of Physics, Graduate School of Science, Osaka University
Machikaneyama 1-1, Toyonaka, Osaka 560-0043, Japan}
\begin{abstract}
The electronic Raman scattering in overdoped (Y,Ca)Ba$_2$Cu$_3$O$_y$ was investigated with changing hole concentration in the superconducting state. It was found that the superconducting responses such as the pair-breaking peaks in the $A_{1g}$ and $B_{1g}$ spectra and the anisotropy of the pair-breaking peak in $XX$ and $YY$ polarizations radically change at 
around the carrier doping $p$=0.19. 
Since both $a$- and  $c$-axis resistivities strongly suggest the closing of pseudogap at $p \sim$ 0.18,
the observed change at $p$=0.19 in superconducting Raman response is attributed to the electronic crossover due to the collapse of the pseudogap. 

\end{abstract}
\maketitle

\section{Introduction}

   A rough sketch of the electronic phase diagram for high-$T_c$ superconducting cuprates (HTSC) was established at the early stage of these twenty years-studies \cite{Torrance}. However, in spite of a tremendous amount of studies, we are still far from the complete understanding of the phase diagram. Compared to the underdoped electronic state governed by a mysterious pseudogap \cite{timusk}, the overdoped state has been less studied so far. It is partly because the electronic state is supposed to merely approach a conventional Fermi liquid metal. 
 
Recently the overdoped state is reexamined in detail, in relation to a quantum critical point (QCP).  Tallon and coworkers have been insisting a QCP at the carrier doping level $p$=0.19 where the pseudogap energy estimated from specific heat falls to zero \cite{tallon}. A recent neutron experiment also suggested that the resonant magnetic modes show a qualitative change at the same doping level \cite{pailhes}. On the other hand, there are reports that physical properties are clearly distinguished between the under- and over-doped regimes at the boundary of optimal doping $p$=0.15-0.16 \cite{marel,gedik,ando}. The former results suggest that the pseudogap line and the "$T_c$ dome" are independent in the phase diagram, which indicates that the pseudogap is irrelevant to the superconductivity pairing and not a precursor of superconductivity. By contrast, the latter facts support the picture in which the pseudogap is a precursor of superconductivity and thus its line merges to the $T
 _c$ dome near the optimum doping, as is predicted by the t-J model \cite{nagaosa,tanamoto} and/or phase fluctuation model\cite{emery}. 

One of the approaches to the phase diagram problem is
to see the electronic change through the superconducting properties such as 
a superconducting gap.
Among various experimental methods,
Raman scattering measurement has an advantage in determining {\bf k}-dependence of gap energy as a bulk property \cite{staufer}
in contrast to surface sensitive probes such as angle-resolved photoemission spectroscopy (ARPES).
Thanks to this advantage,
we have been studying superconducting gaps of YBa$_2$Cu$_3$O$_y$ (Y123) by Raman scattering technique.
In the heavily overdoped sample Y$_{0.88}$Ca$_{0.12}$Ba$_2$Cu$_3$O$_y$, 
we found some anomalies in the pair-breaking peaks, 
and interpreted them as the evidence of $s$-wave component admixture and an increase of chain-plane coupling\cite{masui1,masui2}. 
These phenomena are different not only from the behaviors of optimally doped samples
but also from those of the conventional Fermi liquid-like metal (superconductor)
expected in the overdoped regime.
The next interest is how these anomalies develop with doping,
and how they link to the normal state properties or the phase diagram.

To explore this issue, in the present study, 
we examined a precise doping dependence of Raman scattering spectra between $p$=0.16 and 0.22. 
We also measured $a$- and $c$-axis resistivity to monitor the normal state for the same series of crystals.
When we discuss something related to the doping level in HTSC,
it is very important to collect the data of various physical quantities
for the same series of samples by fixing a material.
Here we have chosen the Y123 system,
and prepared a series of detwinned crystals of Y$_{1-x}$Ca$_{x}$Ba$_2$Cu$_3$O$_y$ (Y/Ca123)
with various $x$ and $y$.
Comparing the spectra for a certain doping level $p$ but different $y$,
we can distinguish the CuO-chain contribution from the response of the CuO$_2$ plane.
It has been uncovered that the spectral changes are not monotonic with $p$ 
but show an abrupt change near $p$=0.19
where the resistivity suggests the closing of pseudogap. 
The possibility for the CuO chain contribution to the observed anomalies was completely ruled out. 

\section{Experiments}

Single crystals of Y/Ca123 were grown by a pulling technique \cite{yamada} for various Ca contents. Rectangular shaped samples were cut from as-grown crystals and detwinned under uniaxial pressure, followed by post-annealing in oxygen atmosphere to adjust oxygen content. Carrier doping level ($p$) was controlled by both Ca content ($x$) and oxygen content ($y$). Ca content was determined by an inductively coupled plasma analysis, while oxygen content was estimated from annealing temperature, using the literature data \cite{fisher} given for each Ca content. Fifteen samples with different combination of $x$ and $y$ were prepared for Raman measurement, and 26 for resistivity measurement in total. 
The value of $x$ varies from 0 to 0.12,
while $y$ is from 6.74 to 6.92.
The carrier concentration $p$ was estimated from the empirical relation between T$_c$ and $p$ \cite{tallon_p}. 

Raman scattering spectra were measured in the pseudo-backscattering configuration for various polarizations with using a triple monochromator equipped with a refrigerator cryostat. 
Although the crystal structure of Y/Ca-123 is orthorhombic, all symmetries refer to a tetragonal D$_{4h}$ point group in the present study. $X$ and $Y$ axis are indexed perpendicular to and along the Cu-O chains, while $X'$ and $Y'$ are rotated by 45 degree from $X$ and $Y$, respectively. In order to extract the electronic Raman response, we fit the spectra in the same way as ref.\cite{limonov}. 

Resistivity was measured by means of a standard four probe technique. 
$\rho_a$ was measured for detwinned samples, while crystals for $\rho_c$-measurements were not detwinned. 
So far, there has been no available result of $\rho_c$ for heavily overdoped Y/Ca123, 
because the measurements for $c$-axis properties are difficult with polycrystalline samples or thin single crystals grown by a flux method. 
It should be mentioned that Y123 single crystals grown by a pulling technique is as long as a few millimeter along $c$-axis, 
which enables us to measure $\rho_c$ accurately.

\section{Results}

\subsection{Raman scattering spectra with $A_{1g}$ and $B_{1g}$ symmetries}

\begin{figure*}[h]
 \begin{center}
   \includegraphics[width=7cm]{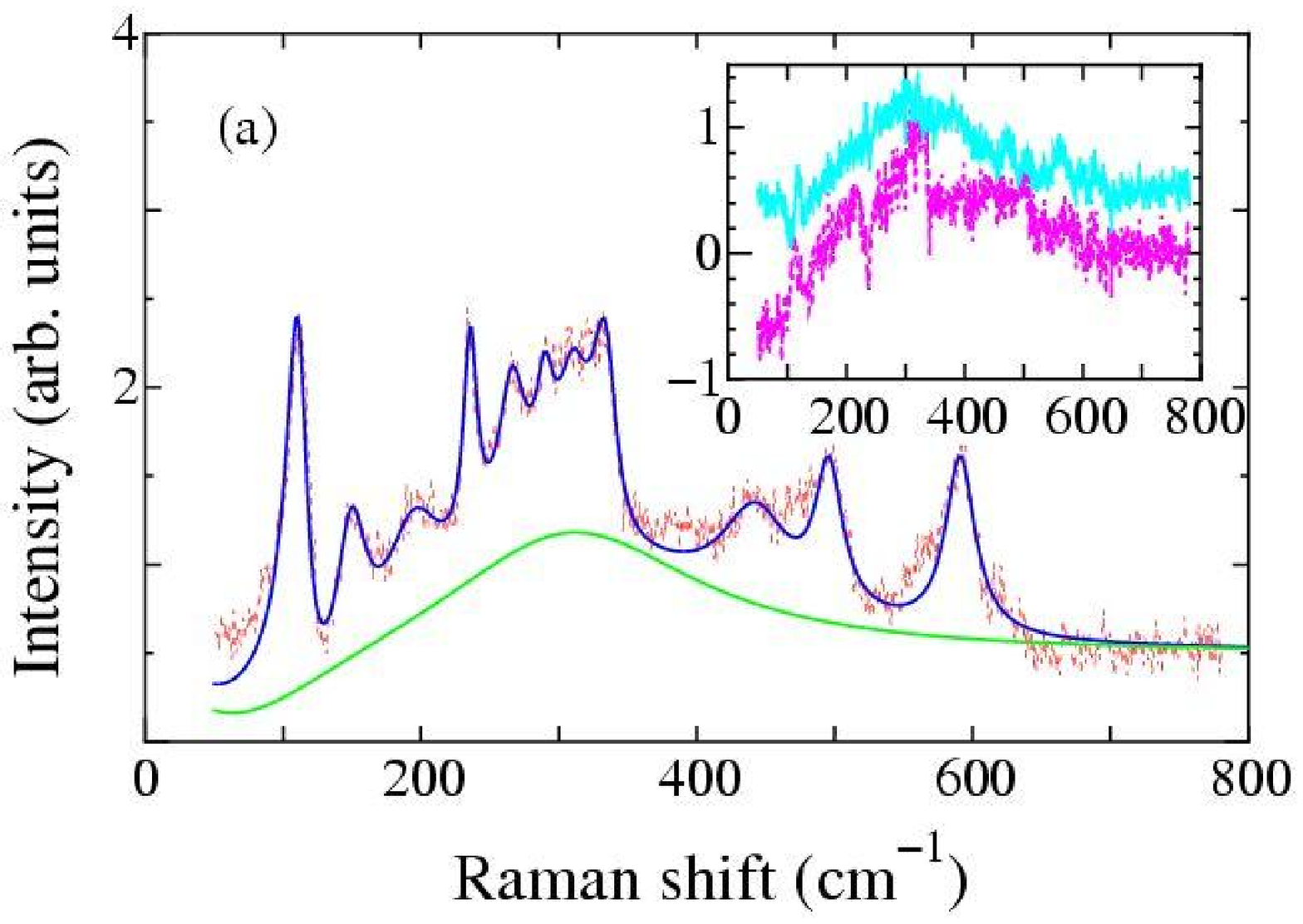} \hspace{5mm}\includegraphics[width=7cm]{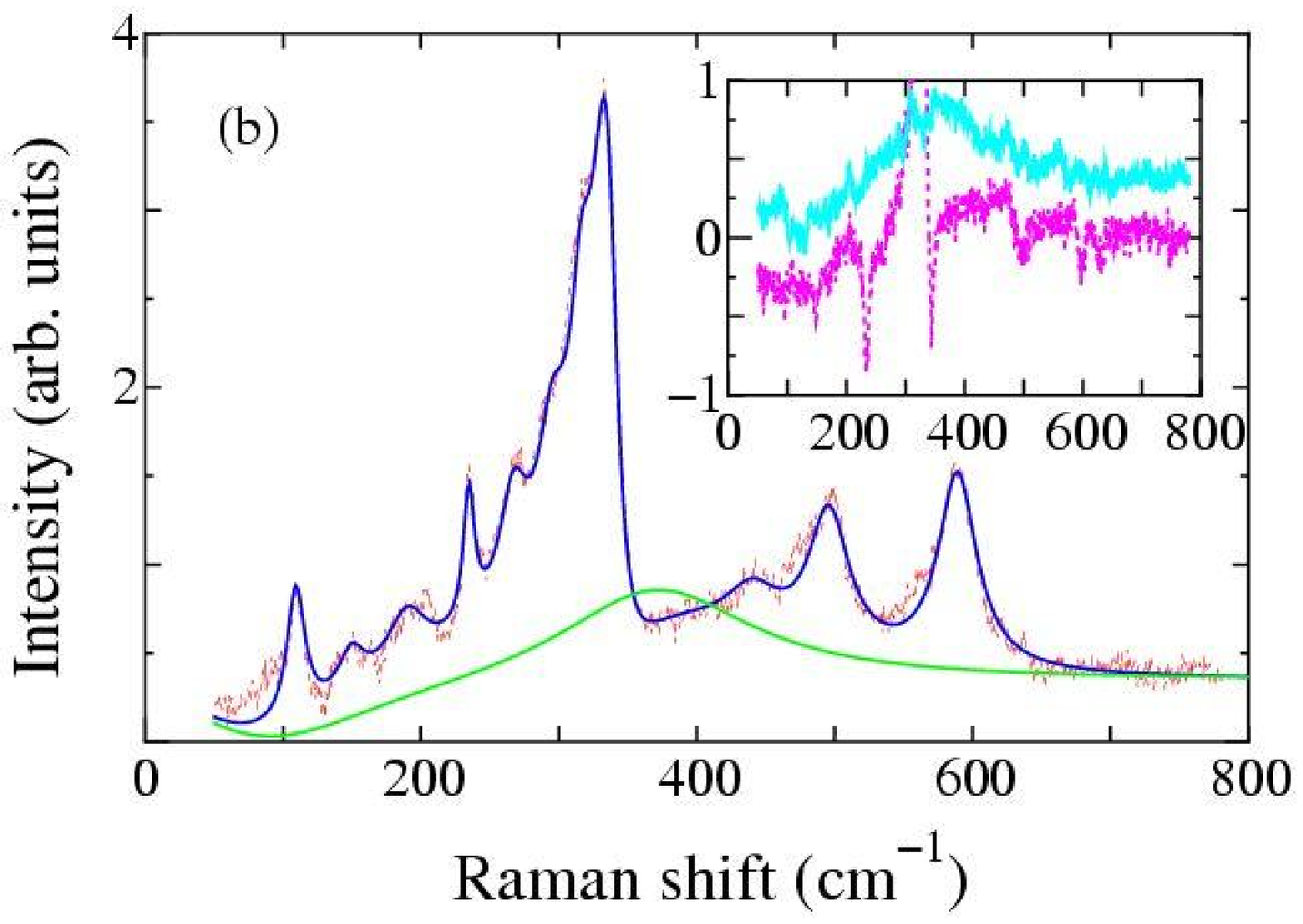}
  \end{center}
\caption{
Raman scattering spectra of Y$_{1-x}$Ca$_x$Ba$_2$Cu$_3$O$_y$ for $x$=0.10, $y$=6.83, and $p$=0.179 with (a) $A_{1g}$ and (b) $B_{1g}$ polarizations. Solid (blue in color) curves are the fitting results by the Green's function method and the light solid (greed in color) is the extracted electronic responses. Insets show electronic response (solid line, cyan in color) and  differences between the spectra at 10K and 100K, I(10K)-I(100K) (dotted line, pink in color). 
}
\label{fitting_spe}
\end{figure*}

\begin{figure*}[h]
 \begin{center}
   \includegraphics[width=7cm]{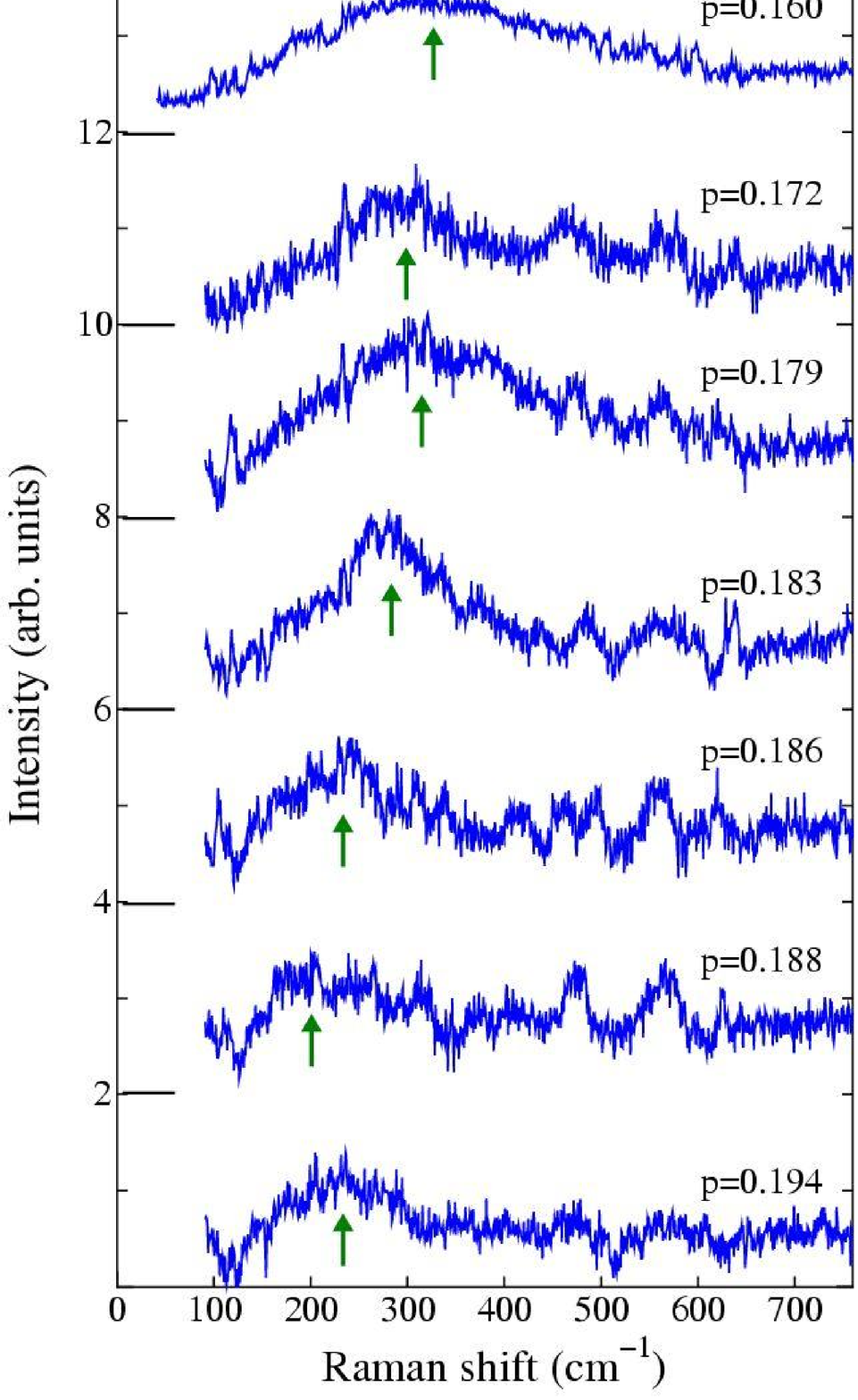} \hspace{5mm}\includegraphics[width=7cm]{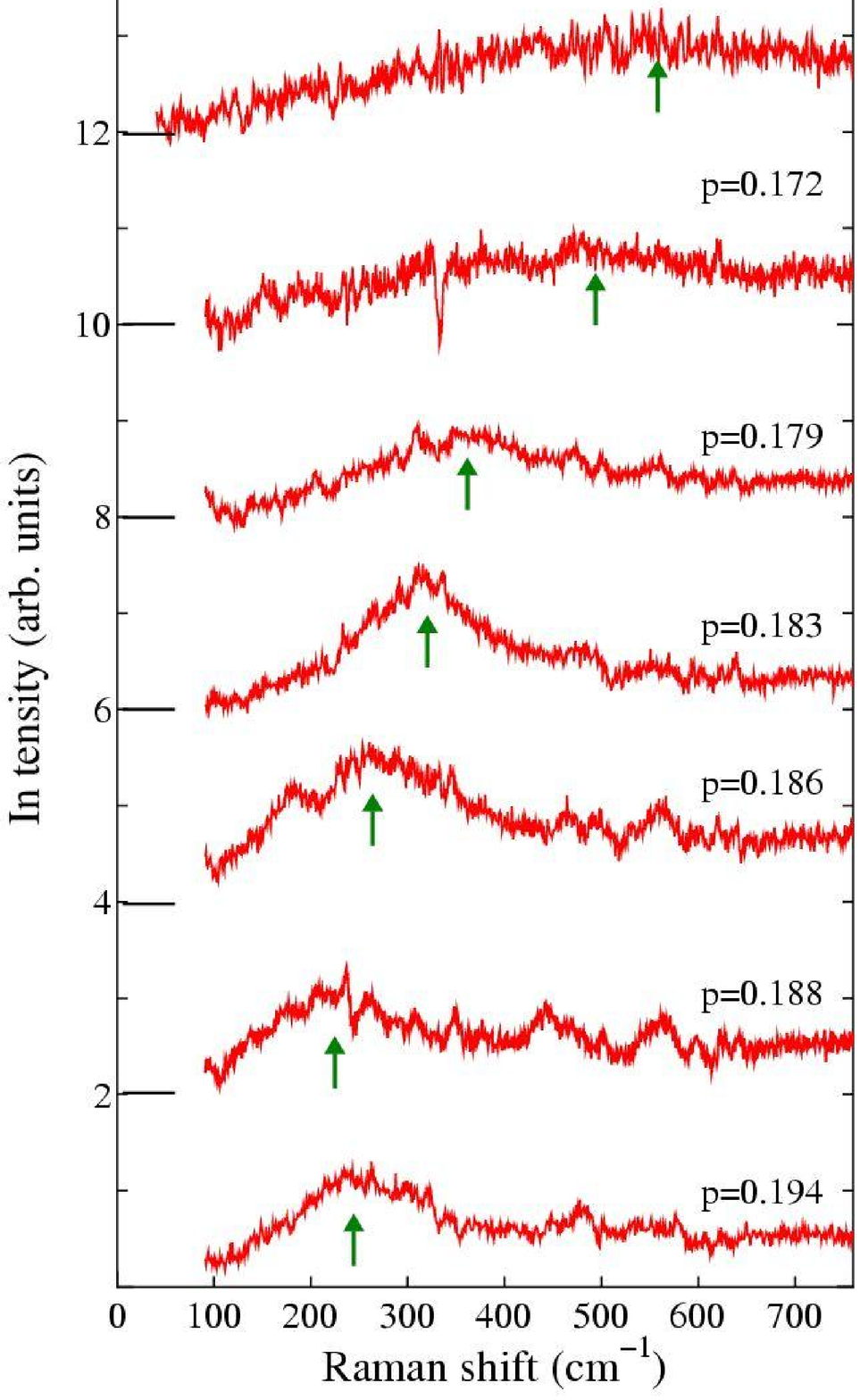}
  \end{center}
\caption{
Doping dependence of the electronic responses in the (a) $A_{1g}$ and (b) $B_{1g}$  Raman scattering (T=10K). Phonon peaks are subtracted by fitting.
}
\label{e-part}
\end{figure*}

    Figure 1 shows the Raman scattering spectra of Y/Ca-123 for $x$=0.103 and $y$=6.83   with $A_{1g}$ ($X'X'-XY$) and $B_{1g}$ ($X'Y'$) polarizations at 10 K. 
Compared to the spectra of optimally doped Y123 ($x$=0 and $y$=6.88)\cite{chen}, 
there appear several sharp peaks ascribed to the phonons induced by oxygen deficiencies. The obtained $A_{1g}$ and $B_{1g}$ electronic Raman spectra (green curves) show broad peaks (pair-breaking peaks) centered at around 300 cm$^{-1}$ and 400 cm$^{-1}$  respectively, which are due to the modification of the electronic component below superconducting transition temperature T$_{\rm c}$. 
The inset of Fig.1 compares the electronic component at 10 K and the difference of the raw spectra at 100K and 10K.  A good correspondence between the two curves justifies our fitting procedure.

   Figure 2 shows gradual changes of the 10K-electronic responses with carrier doping for the $A_{1g}$ and $B_{1g}$ polarizations. In $B_{1g}$ spectra the pair-breaking peak shifts more rapidly to lower energies than in $A_{1g}$ spectra. This decrease is, as expected from the suppression of T$_{\rm c}$, partly due to the decease of superconducting gap energy itself, but is too large to be explained with such a usual mechanism. The difference between $B_{1g}$ and $A_{1g}$ pair-breaking peak energies becomes small with carrier doping. In the electronic Raman scattering of cuprates, the difference between the polarizations is explained by the screening effect, which is usually effective only for $A_{1g}$ polarization \cite{devereaux}. But in the present case, some mechanism are necessary to suppress the peak energy in $B_{1g}$ polarization.   Another remarkable change appears in peak intensity. The intensity of $A_{1g}$ peak decreases with doping, while that of $B_{1g}$ peak increases. 
 All these changes were reported in our previous paper \cite{masui1} and were discussed as the effect of $s$-wave component mixing ($\sim$20 \% at $p$=0.22) into a $d$-wave gap\cite{Nemetschek}.

\begin{figure}[htb]
 \begin{center}
   \includegraphics[width=6.8cm]{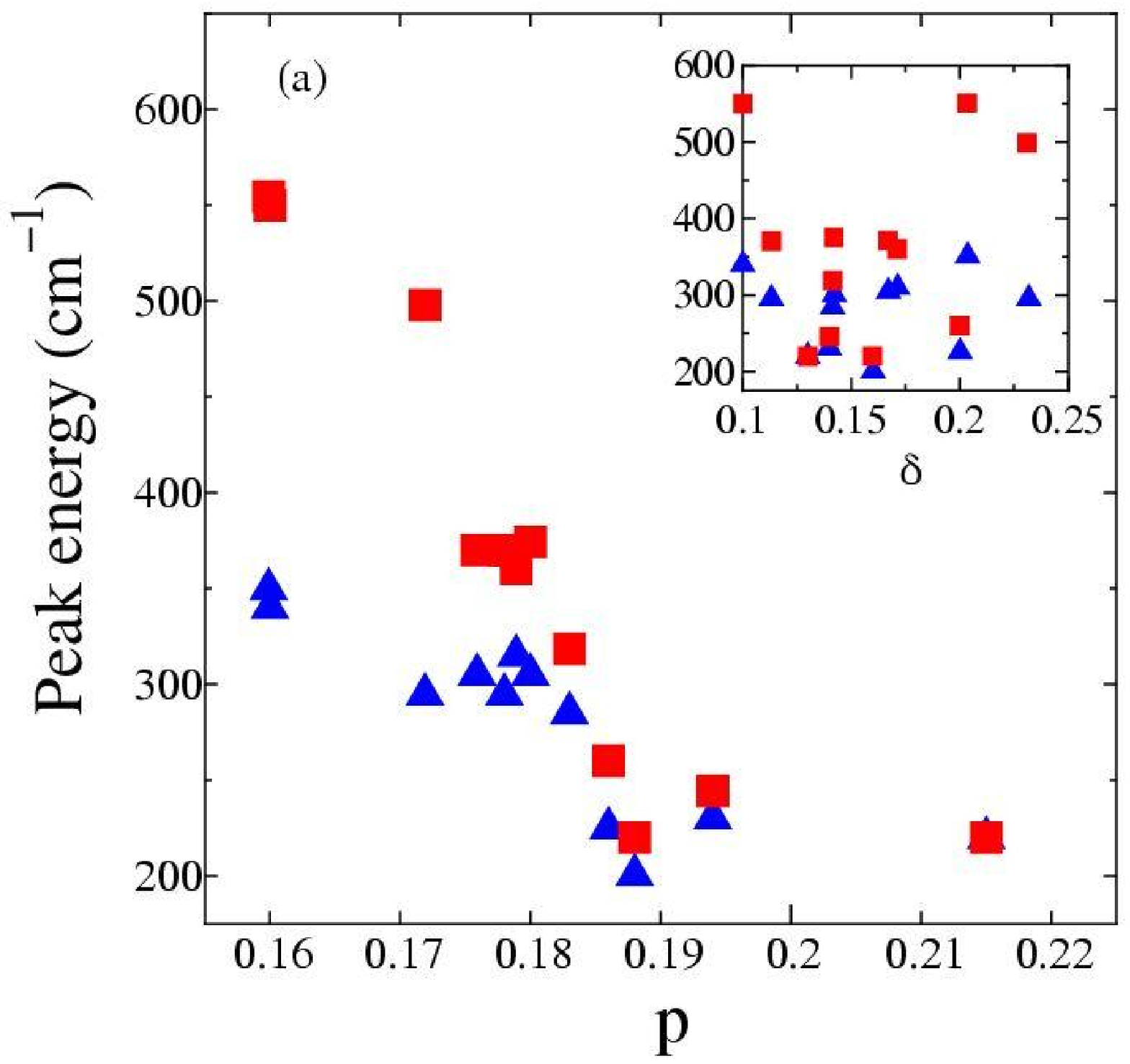}
   \includegraphics[width=6.8cm]{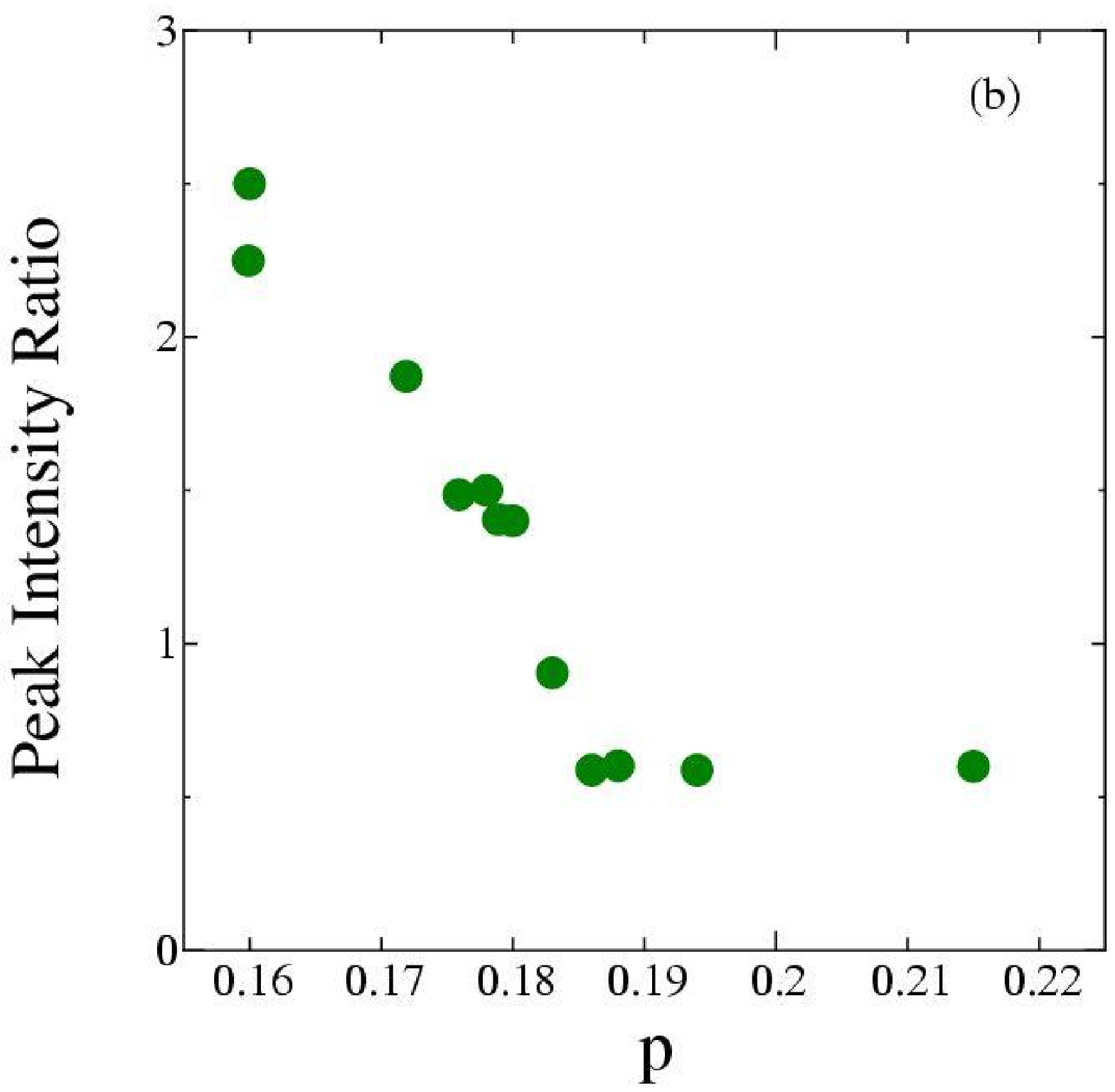}
  \end{center}
\caption{
(a) Doping (p) dependence of the pair-breaking peak energies for $A_{1g}$ and $B_{1g}$ spectra. Inset shows the peak energies as a function of oxygen deficiency $\delta$=7-$y$. Squares and triangles are the data for $B_{1g}$ and $A_{1g}$ peaks, respectively. (b) Doping dependence of the peak intensity ratio I($A_{1g}$)/I($B_{1g}$).
}
\end{figure}
 
 The $B_{1g}$ and $A_{1g}$ peak energies and the peak intensity ratio I($A_{1g}$)/I($B_{1g}$) are plotted in Figures  3(a) and (b). One can find a clear kink at $p\sim$0.19 in the $A_{1g}$ peak energy, while the $B_{1g}$ peak energy rapidly decreases with doping and merges to the $A_{1g}$ line at $p>$0.19.  Here we note that if we plot the same quantities as a function of oxygen deficiency 7$-y$, no systematic change is observed, as demonstrated in the inset. This implies that the observed phenomena are caused only by carrier doping but neither related to the CuO chain structure nor the oxygen deficiency in it. The peak intensity ratio also drops with doping and reaches about 0.5 at $p\sim$0.19, above which it is almost constant. All these results suggest that there is a qualitative difference in the electronic state at $p<$0.19 and $p>$0.19.

\subsection{Raman scattering spectra with $XX$- and $YY$-polarizations}  

Another marked effect of carrier overdoping is the quantum interference between Raman scattering 
of CuO chains and CuO$_2$ planes \cite{masui2}. 
It manifests itself as a strong suppression of the pair-breaking peak in $YY$-polarization spectrum. 
As is typically seen in the Fano line shape for the electron-phonon coupling system, 
the interference between two electronic Raman scattering processes modifies a Raman spectrum, 
if the coupling of these two electronic channels is strong. 
Therefore, the suppression of pair-breaking peak in the $YY$-spectrum is an indication of a 
large transfer matrix between the CuO$_2$-plane and CuO-chain in the heavily overdoped Y/Ca123.

\begin{figure}[htb]
 \begin{center}
   \includegraphics[width=6.8cm]{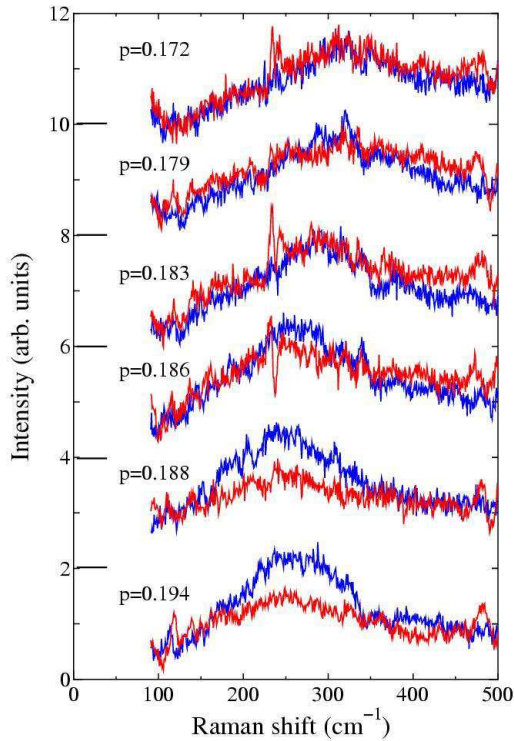}
  \end{center}
\caption{
$XX$- and $YY$-polarized Raman scattering spectra at 10K for various doping levels. The dark (blue in color) and the light (red in color) curves represent the $XX$- and , $YY$-spectra, respectively. Phonon peaks are subtracted by fitting.
}
\end{figure}

 A precise doping dependence of the electronic Raman spectra of $XX$ and $YY$ polarizations are demonstrated in Figure 4. 
It is seen that the suppression of $YY$-polarization peak sets in at around $p$=0.19, 
and then grows with further doping. 
The suppression in $YY$-polarization is not caused by oxygen deficiency in the CuO chain
but develops as a function of doping level.  
Namely, this quantum interference effect is enhanced by the increase of hole concentration on CuO$_2$ planes. 
This is another support for the electronic change at $p\sim$0.19. 
Here we note that the radical shift 
of $B_{1g}$ peak in the tetragonal framework is not an artifact caused by the $XY$-anisotropy. 
It is because the $XY$-anisotropy becomes remarkable at $p>$0.19, 
while the radical shift of $B_{1g}$ peak is observed at $p<$0.19. 
The presence of CuO-chain is essential for the former phenomenon, 
but it plays no role in the latter. 

\subsection{Resistivity in the $a$- and $c$-directions}

\begin{figure}[htb]
 \begin{center}
   \includegraphics[width=6.8cm]{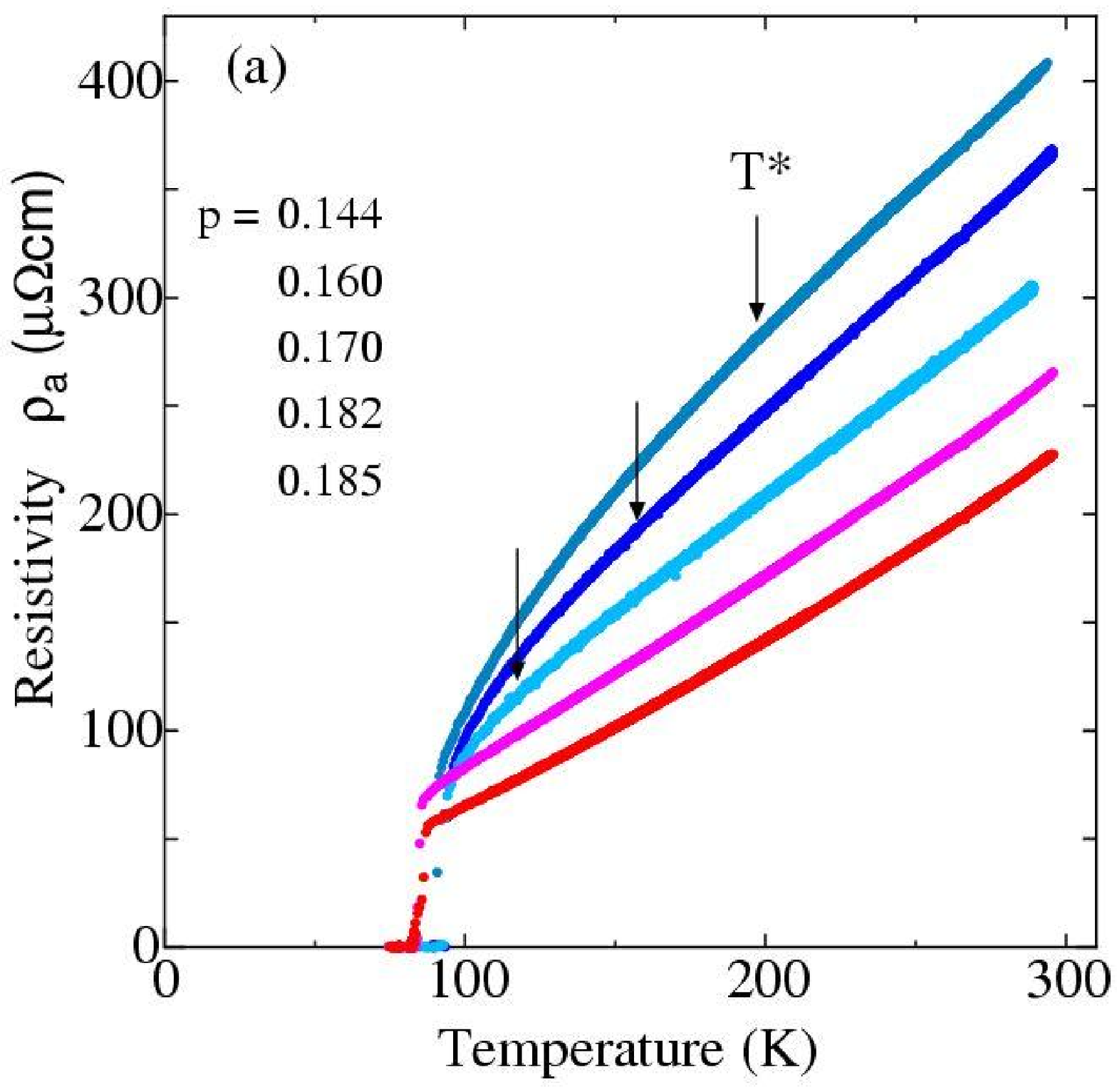}
   \includegraphics[width=6.8cm]{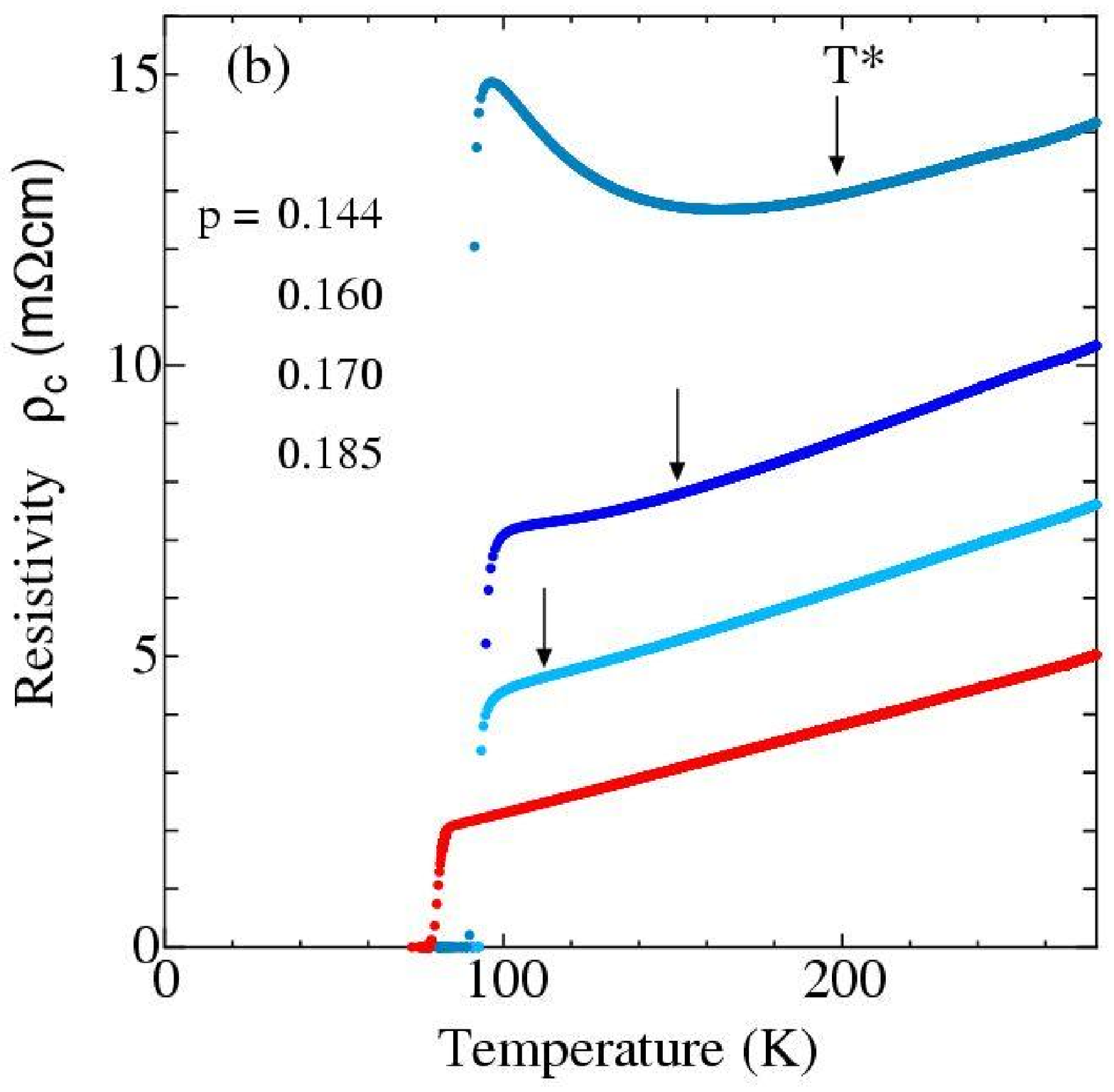}
  \end{center}
\caption{
Resistivities of Y/Ca-123. (a) In-plane resistivities $\rho_a$. Arrows indicate T$^*$ where temperature dependence deviates from T-linear dependence. (b) $c$-axis resistivities $\rho_c$. Arrows indicates T$^*$ where low-temperature upturn of resisvitities starts. 
}
\end{figure}

It is important that the doping level $p\sim$0.19 is a common boundary for the two anomalies in the $A_{1g}$/$B_{1g}$ spectra and the $XX/YY$ spectra. This implies that some distinct change in the electronic state is behind them. To discuss the change at $\sim$0.19, the resistivity data is helpful and suggestive. Figure 5 shows the temperature dependence of resistivity along the $a$- and $c$-axes 
of Y/Ca123 crystals. Since it is a collection of the data for samples with various Ca contents, the doping level does not correlate with oxygen content.  Both $\rho_a$ and $\rho_c$ monotonically decrease with doping. It is rather surprising that $\rho_c$ is not much affected by oxygen deficiency. A signature of pseudogap opening can be seen in the upward deviation of $\rho_c$(T) from the T-linear relation as well as the downward deviation of $\rho_a$(T) from the T-linear relation\cite{takenaka}. 
Figure 5 demonstrates that the pseudogap temperature T$^*$ is lowered with doping and becomes 
invisible above $p\sim$0.18.  
If we extrapolate the T$^*(p)$ line below T$_c$ in the phase diagram, 
it is expected to reach zero at $p\sim$0.19. 
This is consistent with the Tallon's results\cite{tallon}. 

\section{Discussion}

Since a strong two-dimensionality beyond the band theory in HTSC is 
predominantly due to strong electron correlation 
but enhanced by the pseudogap opening,
the close of pseudogap is expected to weaken two dimensionality by increasing 
interlayer coupling. 
On one hand,
the close of pseudogap at $p \sim$ 0.19 is strongly suggested by the disappearance of $T^*$($p$) in Fig.5.
On the other hand,
the negative quantum interference seen in Fig.4 is caused by an increase of transfer matrix 
between CuO chain and CuO$_2$ plane, namely three dimensional coupling between the layers. 
Therefore, it is likely that the negative quantum interference starting at $p$=0.19 is caused by the close of the pseudogap.
It can be considered as the evidence for recovering of three dimensionality above $p$=0.19.

Here it should be noted that even after the pseudogap disappears
anisotropy in the normal state cannot be described by an effective mass model.
This can be seen, for example,
in the temperature dependence of the resistivity ratio $\rho_c/\rho_a$
which should be constant in a Fermi liquid metal.
The value of resistivity ratio ($\sim$ 6 at 90K) is also much larger than the band calculation value.
Therefore, judging from the normal state,
we should say that the electronic state does not return to a three dimensional state in the present doping range.
This seems to be inconsistent with the above statement
based on the Raman response in the superconducting state.
However, it is not obvious whether anisotropy ratios estimated 
from the normal and superconducting state properties
are the same or not.
It is necessary to examine anisotropy in the superconducting state.

In the superconducting state, 
two dimensionality of HTSC is observed in an extremely short coherence length
in the $c$-direction \cite{semba,tomimoto}, 
which results in a large anisotropy ratio $\gamma$ of upper critical fields. 
As we reported in reference \cite{nagasao}, $\gamma$ rapidly decreases with overdoping and 
reaches the band calculation value ($\sim$3) at $p\sim$0.18. 
This is another support for the recovering of three dimensionality observed in the $YY$-spectrum
and for the presence of critical doping at $p$=0.19 where the pseudogap closes.

For the anomalous doping dependence of $A_{1g}$ and $B_{1g}$ pair-breaking peaks, we need more careful discussion.   Four possible factors can be listed up as origins for the decrease in the pair-breaking peak energies: (i) decrease in the maximum gap energy $\Delta_0$ with doping, due to T$_{\rm c}$ suppression, (ii) mixing of the $s$-wave component with doping, (iii) change of the topology of Fermi surface (FS) with doping, (iv) change of the Fermi surface around (0,$\pm\pi$) and ($\pm\pi$,0) by collapse of the pseudogap. It is clear that the first one is not the sole origin to explain the observed rapid decrease of pair-breaking peaks. Some or all of the other three may contribute to this phenomenon. 
The second origin, the $s$-wave mixing is currently the most plausible explanation to reconcile with the observed rapid change of Raman spectra, although the origin of $s$-wave component is unclear yet. As the $s$-wave component increases with doping, the energy of $B_{1g}$ pair-breaking radically decreases because of the appearance of the smaller gap maximum as well as the screening effect in the $B_{1g}$ channel\cite{Nemetschek}. The screening effect also suppresses the $A_{1g}$ peak intensity.

The third one, the FS topology should also be taken into account, 
because the {\bf k}-dependence of superconducting gap energy depends on the shape of FS even for the case of simple $d$-wave gap. 
Moreover, the $B_{1g}$ Raman vertex is quite sensitive to the FS topology in the antinodal direction 
where van Hove singularity is present. 
According to the calculation by Branch and Carbotte, 
the $B_{1g}$ peak energy changes with the second nearest hopping $t'$ because the FS topology changes 
with $t'$ from hole-like to electron-like one \cite{branch}. 
The change of hole-like FS to electron-like one was reported by ARPES 
for La$_{2-x}$Sr$_x$CuO$_4$ with $x\sim$0.22 \cite{ino}. 
Although so far there is no information about the FS topology of overdoped YBCO, 
a similar FS change, if it exists, can be a part of the sources for the observed anomalies in Raman spectra.  
However, the estimated reduction of $B_{1g}$ peak energy is 30 \% at most 
by the change of FS topology. 

Since the $B_{1g}$ Raman spectrum is sensitive to the FS around ($\pm\pi$,0) and (0,$\pm\pi$), 
disappearance of the pseudogap (the fourth factor) must seriously affect the $B_{1g}$ spectrum. 
The next question is which of these four factors can cause the non-monotonic change 
in the $A_{1g}$ and $B_{1g}$ pair-breaking peaks.  
The first factor (decrease in $\Delta_0$) can be ruled out, 
because it is unlikely that $\Delta_0$ abruptly changes at $p$=0.19.
The observed spectral change at p$\sim$0.19 should be rather
attributed to the FS change in the antinodal direction,
in particular, to closing of the pseudogap. 

Similar anomalous changes in Raman spectra at a certain critical doping were also observed 
for Tl$_2$Ba$_2$CuO$_z$ \cite{kendziora,gasparov,nishikawa}, 
and thus this should be common among HTSC.
Although one may expect that the collapse of pseudogap in overdoped regime leads to a simple $d$-wave superconductivity, the observed anomaly in this study is quite distinct from such a simple superconducting state. Here we may take into consideration the existence of unpaired carriers in the overdoped regime. For example, a large amount of unpaired carriers are observed in far-infrared spectra\cite{schtzmann}, specific heat\cite{loram}, and magnetic susceptibility\cite{tanabe}. If a superconducting phase and unpaired carriers coexist, a proximity effect between the two phases must be taken into account. This could induce an anomaly such as an $s$-wave component. An abrupt change at $p$=0.19 suggests that this proximity effect may change where the FS is fully recovered. 

\section{Summary} 

In order to examine the phase diagram from the viewpoint of superconducting response,
we studied the electronic Raman spectra of overdoped (Y,Ca)Ba$_2$Cu$_3$O$_y$
as a function of doping level $p$. 
By a precise and systematic study on a series of detwinned crystals with various Ca and oxygen contents,
we were able to remove the contribution of oxygen deficiency in our discussion
and to extract a pure doping effect.

It was found that the changes of superconducting responses in $XX/YY$- and $A_{1g}/B_{1g}$-polarizations are not
monotonic but show abrupt changes at around $p$=0.19, 
where the closing of pseudogap is suggested by both of $\rho_a$(T) and $\rho_c$(T). 
These changes at $p$=0.19 are attributed to an essential change of the electronic state 
such as the closing of pseudogap and the change of Fermi surface topology. 
Our results support the picture in which the pseudogap line T$^*$($p$) hits to zero at $p\sim$0.19 
in the phase diagram for Y123. 
The negative quantum interference in Raman scattering and the anisotropy ratio of
upper critical fields suggest a three dimensional electronic state above $p$=0.19,
while the normal state resistivity indicates a persistence of unusual charge dynamics
along the $c$-axis.

This work is supported by New Energy and Industrial Technology Development Organization (NEDO) as Collaborate Research and Development of Fundamental
Technologies for Superconductivity Applications.

\end{document}